\documentstyle[psfig,times]{jaa}
\psfigurepath{dir:/home/sushant/CME/PRL/}
\begin{document}
%\title{HELIOSEISMIC RING ANALYSIS OF CME SOURCE REGIONS}
\title{Helioseismic ring analysis of CME source regions}
\author[S.C. Tripathy {\it et al.}]%
{S. C. Tripathy$^1$\thanks{e-mail: stripathy@nso.edu},
S. de Wet$^{1,2}$,  K. Jain$^1$, R. Clark$^1$, and  F. Hill$^1$ \\
$^1$ National Solar Observatory
Tucson, AZ 85719, USA. \\
$^2$ Rice University, Houston, TX, 77005, USA.}
%\pubyear{1995}
%\volume{16}
%\setcounter{page}{217}
%\date{Received 1996 June 20; accepted 2000 March 09}
\maketitle
\label{firstpage}
\begin{abstract}
 We apply the
ring diagram technique to  source regions of halo coronal mass ejections (CMEs)
to study changes in acoustic mode parameters
  before, during, and after the onset
of CMEs.    We find that CME 
regions associated with a low value of magnetic flux have line widths smaller  than the quiet regions
implying a longer life-time  for the 
oscillation modes.   We suggest that this criterion may be used to forecast the active regions which may 
trigger CMEs. 
\end{abstract}

\begin{keywords}
Sun -- coronal mass ejections -- oscillations  -- magnetic field
\end{keywords}
\section{Introduction}\label{sec:intro}

Using high-resolution Global Oscillation Network Group (GONG) Dopplergrams, we
have analysed the changes in oscillation mode properties in the source locations of 
CMEs. This was undertaken to find out if the mode properties of the CME regions 
 are different from the active regions which did not 
produce a CME. This may help to identify the regions on solar disk which may produce  
CMEs.   

\section{Data and Analysis }\label{sec:data}
The location of the CMEs are taken from the catalogue of Zhou {\it et al}. (2006) 
who have identified the source regions of 288 halo CMEs during the period of 1997-2003. 
This data base lists the location, date, time of occurrence and 
the classification of the region where the CMEs occured.     
To infer the mode parameters,  we use the ring-diagram analysis (Hill, 1988) and 
 compute power spectra of solar 
oscillations over  smaller patches of the solar surface corresponding to the CME regions. 
The mode parameters are then 
extracted by fitting  the power spectrum with a
symmetric profile model  which considers the peaks in the power
spectra to be Lorentzian (Haber {\it et al}., 2001).
\section{Results and Discussion}
 In this analysis we examine source locations of
48 CMEs during September 2001 -- December 2003 that occurred no more than 30 degrees in any of the four 
cardinal directions from the solar disk center. These  are further associated with flaring, active, 
and filament regions. % (Tripathy {\it et al.} 2007).  
The power spectra of these CME regions were calculated from an area covering 128 $\times$ 128 pixels 
with the center coinciding with the position of the CME. The 
area roughly corresponds to about 15 $\times$ 15 degrees in heliographic longitude and latitude.  
Each region was tracked for 1664 minutes where the central time corresponds to 
the onset time of the CME. In order to estimate the variation in mode parameters, we compare 
each of the CME region with 
a magnetically quiet region at the same latitude as the CME region within the same Carrington rotation.
The average magnetic flux ($B_{av}$) associated with each CME and quiet region is calculated from the 
synoptic maps of Kitt Peak magnetograms (ftp://nsokp.nso.edu/kpvt/synoptic/mag/)  
by averaging the unsigned magnetic flux inside the 
same 15 $\times$ 15 degree box of the corresponding regions.
\begin{figure}[t]          %%%%FIG 1
%\vspace{-0.2in}
\centering
\psfig{file=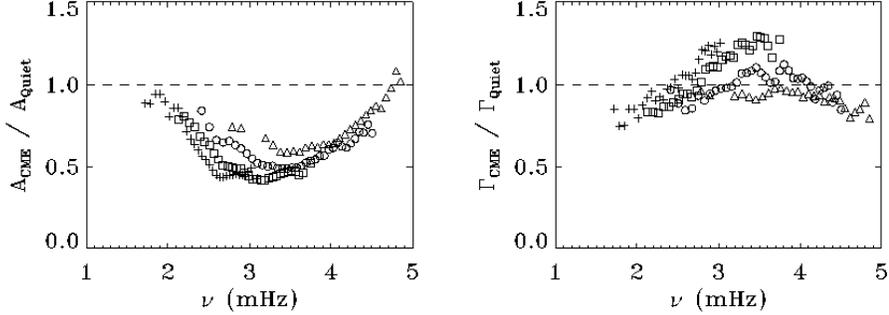,angle=90,width=12cm,clip=}
\vspace{-0.2in}
\caption{Changes in (a) ratio of the peak power 
and (b) ratio of the half-width  between the CME region of  December 19, 2002 ($B_{av}$ = 113 G) 
and quiet region of December 12, 2002 ($B_{av}$ = 3 G) both located at N22W12.  The {\it f}-modes 
are represented by + and {\it p}-modes are 
represented by diamonds, circles, and triangles for $p_1$ through $p_3$ respectively.
 \label{fig1}}
\end{figure} 

 As an example of the variation in mode parameters in CME regions, we have shown the
 ratio of the peak power ($A$) between a CME and a quiet region in Figure~1a. 
It is evident that {\it f} and {\it p}-modes in the CME region 
have lower power than the quiet region and the variation is not a monotonic function of the frequency. 
The maximum suppression seems to occur in the frequency range of about 3-3.5 mHz 
and the suppression decreases with increasing radial order. 
At high frequencies, the suppression tends to become small and in few cases, 
we even note enhancement of power. In general, the results are 
consistent with those related to active regions (Rajaguru, Basu, and Antia, 2001).

The variation in half-width ($\Gamma$) between the same pair of CME and quiet region is shown in  Figure~1b
as the ratio between the two.   
For  {\it f} and {\it $p_1$} modes, the half-widths are larger in higher end of the frequency range and smaller 
in lower end of the frequency range. The widths of other {\it p}-modes are generally smaller than 
those in quiet regions.  Since, the lifetime of a mode is inversely proportional to its 
half-width, the small width indicates that  some modes live longer in regions where CMEs originate
and implies a slow damping process. This result is in contrast to the study of active regions  
(Rajaguru, Basu, and Antia, 2001) where the 
lifetime of the modes were found to be smaller compared to the quiet regions. 

\begin{figure}           %%%%FIG 2
\centering
\hskip 0.1in
\psfig{file=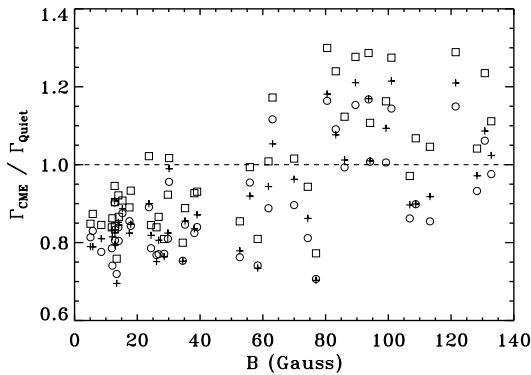,angle=90,width=8cm,clip=}
\vskip -0.2in
\caption{Frequency averaged ratio of half-widths for 48 CME regions as a function of magnetic flux.  
The symbols have the same meaning as in Figure~1. 
 \label{fig2}}
\end{figure}

Figure 2 shows the frequency averaged  ratios of the half-width between the CME and quiet region for 
48 CMEs as a 
function of average magnetic flux of the CME region. The widths are averaged over the frequency range of 
2550--2750 $\mu$Hz for {\it f}-modes and 3000--3500 $\mu$Hz for {\it p}-modes. 
It is evident that the line width 
depends on the strength of magnetic 
field present in the region. Most of the CMEs 
associated with low  magnetic flux have a smaller width compared to the quiet region. 
The width increases for higher magnetic flux values. However, there are exceptions in both the cases. 

\begin{figure}[t]           %%%%FIG 3a
\vspace{-1.5in}
\centering
\psfig{file=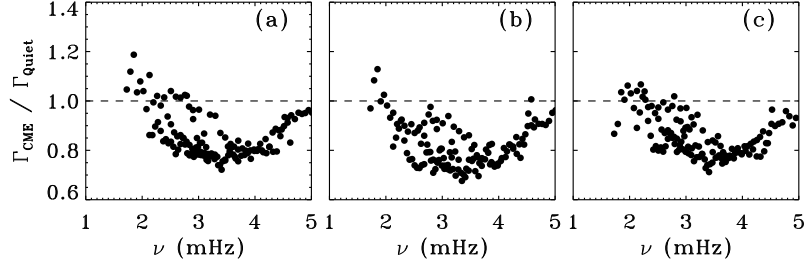,angle=90,width=12cm,clip=}
\vskip -0.25in
\caption{The changes in half-width as a ratio between a CME and  
quiet region for three consecutive days.  Panel (b) refers to the day of onset of the CME 
while panels (a) and (c)
 refers to one day before and after the CME occured, respectively.       
 \label{fig3a}}
\end{figure}
We have also studied the evolution of the oscillation mode parameters of the CME region one day prior 
and one day after the onset of the CME. As an example, we show the event of March 17, 2002 where the CME 
is located at W10S0 and  has an average  flux of 14 G (Figure~3). The quiet region for this event corresponds 
to the same position 
on April 4, 2002 and has a magnetic flux of 6.4 G.  
  To compare, we also show the changes in half-width  for three consecutive days 
for the active region NOAA 9628  
located at S22.5E15 on September 23, 2001 (Figure~4). 
On comparing Figures~3 and 4, it is evident that in
the region where the CME is triggered, the line width is always less than the quiet region 
while in case of active region, it is always higher than the quiet region.  We propose 
this characteristic property of the oscillation modes as the prime signal to identify an active 
region which may produce a CME.

\begin{figure}[h]  
\vspace{-1.7in}         %%%%FIG 3b
\centering
\psfig{file=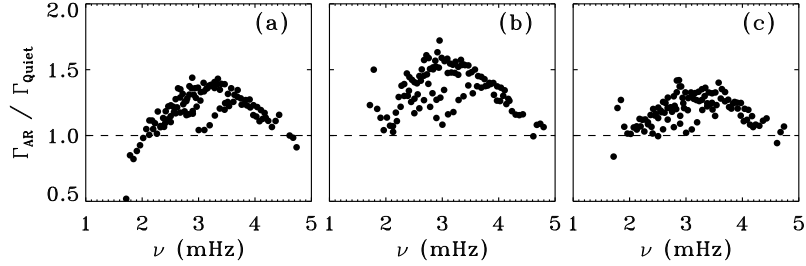,angle=90,width=12cm,clip=}
\vskip -0.4in
\caption{Same as Figure~2 but for the active region NOAA 9628.  \label{fig3b}}
\end{figure}

In summary, we find that the oscillation modes  in the source 
locations of CMEs with a low value of magnetic flux have a higher life time as compared to the quiet 
regions. This indicates that damping process of the modes associated with CME regions 
are slow as compared to the quiet regions.

%\acknowledgements
A part of this work is carried out through the National Solar Observatory 
Research Experiences for Undergraduate (REU) site program.
This work utilizes data obtained by the Global Oscillation Network
Group program, managed by the National Solar Observatory, which
is operated by AURA, Inc. under a cooperative agreement with the
National Science Foundation. The data were acquired by instruments
operated by the Big Bear Solar Observatory, High Altitude Observatory,
Learmonth Solar Observatory, Udaipur Solar Observatory, Instituto de
Astrof\'{\i}sica de Canarias, and Cerro Tololo Interamerican
Observatory. NSO/Kitt Peak data used here are produced cooperatively by
NSF/NOAO, NASA/GSFC, and NOAA/SEL. This work is supported by NASA grant NNG 5-11703 and NNG 05HL41I.

\label{lastpage}
\end{document}